# The postperovskite transition in Fe- and Al-bearing bridgmanite: effects on seismic observables

Juan J. Valencia-Cardona[1], Renata M. Wentzcovitch[2,3,4*], Jingyi Zhuang[3,4], Gaurav Shukla[5], Kanchan Sarkar[1,4]

[1] University of Minnesota, Department of Chemical Engineering and Material Science, Minneapolis, 55455, USA.

[2] Columbia University, Department of Applied Physics and Applied Mathematics, New York City, 10027, USA.

[3] Department of Earth and Environmental Sciences, Columbia University, New York, NY, 10027, USA.

[4] Lamont-Doherty Earth Observatory, Columbia University, Palisades, 10964, USA.

[5] Indian Institute of Science Education and Research Kolkata, Department of Earth Sciences, Mohanpur, West Bengal, India.

*Corresponding author: Renata M. Wentzcovitch (rmw2150@columbia.edu).

**Key Points:**

- *Ab initio* results on the Pv-PPv transition in Fe- and Al-bearing bridgmanite are consistent with experimental measurements in phases with similar compositions.

- Plausible compositions display wide transition pressure ranges. A change of aggregate composition across a sharp D" discontinuity is likely to occur.

- The Pv-PPv exothermic transition does not contribute significantly to the thermal boundary layer atop the CMB.

**Abstract**

The primary phase of the Earth's lower mantle, (Al, Fe)-bearing bridgmanite, transitions to the postperovskite (PPv) phase at Earth's deep mantle conditions. Despite extensive experimental and *ab initio* investigations, there are still important aspects of this transformation that need clarification. Here, we address this transition in ($Al^{3+}$, $Fe^{3+}$)-, ($Al^{3+}$)-, ($Fe^{2+}$)-, and ($Fe^{3+}$)-bearing bridgmanite using *ab initio* calculations and validate our results against experiments on similar compositions. Consistent with experiments, our results show that the onset transition pressure and the width of the two-phase region depend distinctly on the chemical composition: a) $Fe^{3+}$-, $Al^{3+}$-, or ($Al^{3+}$, $Fe^{3+}$)-alloying increases the transition pressure, while $Fe^{2+}$-alloying has the opposite effect; b) in the absence of coexisting phases, the pressure-depth range of the Pv-PPv transition seems quite broad to cause a sharp D" discontinuity (< 30 km); c) the average Clapeyron slope of the two-phase regions are consistent with previous measurements, calculations in $MgSiO_3$, and inferences from seismic data. In addition, d) we observe a softening of the bulk modulus in the two-phase region. The consistency between our results and experiments gives us the confidence to proceed and examine this transition in aggregates with different compositions computationally, which will be fundamental for resolving the most likely chemical composition of the D" region by analyses of tomographic images.

**1 Introduction**

The extraordinary complexity of the D" region, i.e., the layer ∼ 300 km above the core-mantle boundary (CMB), originates in this region's enigmatic composition and thermal structure. In particular, the thermal boundary layer is essential for understanding the heat flux across the CMB and its implications for the thermal evolution of the core and geodynamo generation (Nakagawa & Tackley, 2004). Phase transitions in this region can produce discontinuities in tomographic models or regional one-dimensional seismic models (Lay et al., 2004; Wysession et al., 1998). The D" velocity discontinuity varies laterally in depth, width, and magnitude. Shear velocity ($V_S$) discontinuity varies between ∼ 2.5% and 3% in over 40 studies summarized in Wysession *et al*. (Wysession et al., 1998). Compressional velocity ($V_P$) discontinuity is systematically smaller but varies more widely (Wysession et al., 1998). Synthetic waveforms computed from geodynamic simulations suggested that this discontinuity pattern was most consistent with a thermal slab interacting with a phase transition (Sidorin et al., 1998).

In 2004, it was found that the major mineral phase of the lower mantle, $MgSiO_3$ perovskite (Pv), or bridgmanite (Bdm), transforms into the seemingly layered postperovskite (PPv) phase at approximately 125 GPa and 2,500 K depth (Murakami et al., 2004; T. Tsuchiya et al., 2004; Oganov & Ono, 2004). Since then, this transition has become a top candidate for the origin of the D" discontinuity. Following this discovery, extensive experimental and computational (e.g., Iitaka et al., 2004; Shim et al., 2004; Mao et al., 2004; Akber-Knutson et al., 2005; Caracas & Cohen, 2005; Tateno, 2005; Ono & Oganov, 2005; Wentzcovitch et al., 2006; Tateno et al., 2007; Murakami et al., 2007; Sinmyo et al., 2008; J. Tsuchiya & Tsuchiya, 2008; Hirose et al., 2008; Catalli et al., 2009; Andrault et al., 2010; Grocholski et al., 2012; Metsue & Tsuchiya, 2012; Wang et al., 2019; Kuwayama et al., 2022) studies have been carried out. Despite all these efforts, essential aspects of this transformation still need clarification, particularly the effect of alloying elements such as Fe and Al. These elements' presence or absence changes the Pv-PPv transition pressure and opens a two-phase coexistence region with a chemistry-dependent pressure range and Clapeyron slopes. For instance, various studies have argued that alloying with $Al_2O_3$ increases the transition pressure (Akber-Knutson et al., 2005; Caracas & Cohen, 2005; Grocholski et al., 2012; Tateno, 2005), while others observed a moderate decrease in transition pressure (J. Tsuchiya & Tsuchiya, 2008; Wang et al., 2019). Similar disagreement exists with ferrous iron ($Fe^{2+}$) alloying: some studies have found that $Fe^{2+}$-Bdm lowers the Pv-PPv transition pressure (Caracas & Cohen, 2005; Mao et al., 2004; Ono & Oganov, 2005; Shieh et al., 2006; Sun et al., 2018), while others (Hirose et al., 2008; Tateno et al., 2007) have found the opposite. Furthermore, Catalli et al. (Catalli et al., 2009) showed that the ferric iron ($Fe^{3+}$) bearing phase had a somewhat narrow Pv-PPv coexistence pressure range while simultaneous ($Fe^{3+}$, $Al^{3+}$) substitution had a coexistence range similar to that displayed by the $Fe^{2+}$- bearing system. However, both $Fe^{3+}$- and ($Fe^{3+}$, $Al^{3+}$)-bearing systems increased the onset transition pressure.

Seismic velocities reduction at the bottom of the D" region has also been detected (Thomas et al., 2004; Wysession et al., 1998). Some of these velocity reduction patterns have been interpreted by Hernlund (Hernlund, 2010; Hernlund et al., 2005; Hernlund & Labrosse, 2007) as a "double-crossing" phenomenon, where Pv transforms to PPv and back-transforms to Pv with increasing depth. This phenomenon depends strongly on the phase boundary but also on the local geotherm: lower temperatures stabilize the PPv phase, steep temperature gradients enhance the chances of double-crossing, and very steep temperature gradients may prevent the occurrence of the Pv-PPv transition (Hernlund et al., 2005). The seismic signature of the Pv-PPv transition also depends critically on the rock composition (Grocholski et al., 2012; Kuwayama et al., 2022), and this issue also needs more clarification. In particular, the transition pressure range in a pyrolytic aggregate is still debated.

Here we use *ab initio* thermodynamic results (Shukla et al., 2015, 2016, 2019) on these Pv and PPv alloys to compute the two-phase boundary regions and test them against experiments. This effort is a fundamental step for methodology validation before computing the phase boundary in aggregates. We address the width of the coexistence regions, Clapeyron slope, and the effect on seismic velocities along two geotherms, an adiabatic one (Brown & Shankland, 1981) and another containing a thermal boundary layer above the CMB (Boehler et al., 2008). We also compute the isentropes throughout the Pv-PPv transition to address the potential effect of this transition on the thermal boundary layer. The Pv-PPv transition in aggregates in thermochemical equilibrium is beyond the scope of this work. This step will be taken in follow-up work.

## 2 Methods and Calculation Details

### 2.1 *Ab initio* calculations

*Ab initio* results of static, vibrational, and elastic properties of bridgmanite (Pv) and postperovskite (PPv) phases of $Mg_{1-x}Fe^{2+}_{x}SiO_3$, $(Mg_{1-x}Fe^{3+}_{x})(Si_{1-x}Fe^{3+}_{x})O_3$, $(Mg_{1-x}Al_{x})(Si_{1-x}Al_{x})O_3$ and $(Mg_{1-x}Fe^{3+}_{x})(Si_{1-x}Al_{x})O_3$ with the bearing element concentration $x = 0$ and 0.125 are from Shukla et al. (Shukla et al., 2015, 2016, 2019). These DFT+$U_{SC}$ (LDA+$U_{SC}$ and GGA + $U_{SC}$) calculations used the Quantum ESPRESSO software (Giannozzi et al., 2009, 2017) and were performed on supercells with up to 80 atoms. More computational details are reported in those studies. $Fe^{2+}$ and ($Fe^{3+}$, Al) in Pv and PPv are in the high-spin (HS) state, while $Fe^{3+}$ in the A-site B-site is in the LS state (Hsu et al., 2010, 2011, 2012; Yu et al., 2012). The HS to LS state change in iron in these systems occurs at pressures considerably lower than the Pv to PPv transition (Hsu et al., 2011; Yu et al., 2012). Thermodynamic and thermoelastic properties are here recalculated using the quasi-harmonic approximation (Wallace, 1972) using the `qha` (Qin *et al.*, 2019) and `cij` codes (Luo et al., 2021). The performance of these approaches for thermodynamic and thermoelastic properties have been extensively documented (Wentzcovitch, Wu, et al., 2010; Wentzcovitch, Yu, et al., 2010).

## 2.2 Thermodynamic modeling

Free energies were computed for Pv and PPv alloys using a "quasi-ideal" solid solution model and the standard state model for binary systems (Gaskell, 2008), where the Gibbs free energy of mixing curves of the Pv and PPv phases in an A-B binary system, at a given as (see Figure S1):

$$\Delta G_{Pv}^M = k_B T(X_B \ln[X_B] + (1 - X_B) \ln[1 - X_B]) + X_B(\Delta G_B^{Pv-PPv}) \quad (1)$$

$$\Delta G_{PPv}^M = k_B T(X_B \ln[X_B] + (1 - X_B) \ln[1 - X_B]) + (1 - X_B)(\Delta G_A^{PPv-Pv}) \quad (2)$$

with $k_B$ as the Boltzmann constant. End-member' A' is $MgSiO_3$, while end-member 'B' is either $Mg_{0.875}Fe^{2+}_{0.125}SiO_3$, or $(Mg_{0.875}Fe^{3+}_{0.125})(Si_{0.875}Fe^{3+}_{0.125})O_3$, or $(Mg_{0.875}Al_{0.125})(Si_{0.875}Al_{0.125})O_3$, or $(Mg_{0.875}Fe^{3+}_{0.125})(Si_{0.875}Al_{0.125})O_3$. In this case, $X_B = x/0.125$, where x as the alloying element concentration. Since end-member B is not a pure phase, its partial molar Gibbs free energy includes non-ideality, i.e., the interaction between, e.g., $MgSiO_3$ and $FeSiO_3$ end-members, and configuration entropy. Therefore, although treated as ideal, the A-B system captures most of the excess Gibbs free energy in the B end-member free energy. We refer to this solution as "quasi-ideal." It corresponds to a Henryan solution modeling of an A-B alloy with end-member B being $FeSiO_3$, $Fe_2O_3$, $Al_2O_3$, or $FeAlO_3$ in the low concentration limit and having a constant activity coefficient but different from 1.0. In the expressions above, $\Delta G_A^{Pv-PPv} = G_A^{Pv} - G_A^{PPv}$ and, $\Delta G_B^{Pv-PPv} = G_B^{Pv} - G_B^{PPv}$. The compositions of the Pv and PPv solvus lines, $X_B^{Pv}$ and $X_B^{PPv}$, are given by (Gaskell, 2008):

$$X_B^{PPv} = \frac{1 - e^{\frac{G_A^{PPv} - G_A^{Pv}}{k_B T}}}{e^{\frac{G_B^{PPv} - G_B^{Pv}}{k_B T}} - e^{\frac{G_A^{PPv} - G_A^{Pv}}{k_B T}}}, \quad X_B^{Pv} = X_B^{PPv}\left(e^{\frac{G_B^{PPv} - G_B^{Pv}}{k_B T}}\right) \quad (3)$$

The PPv fraction, $n_{PPv}$, is given by the lever rule:

$$n_{PPv} = \frac{X_B - X_B^{Pv}}{X_B^{PPv} - X_B^{Pv}} \quad (4)$$

The isentropes of the two-phase aggregate are computed by integrating the adiabatic gradient

$$\left(\frac{\partial T}{\partial P}\right)_S = \frac{\alpha V T}{C_P} \quad (5)$$

where $\alpha$, $V$ and $C_P$ are the thermal expansion coefficient, volume, and isobaric specific heat of the two-phase aggregate. These aggregate properties are given by $V = n_{Pv}V_{Pv} + n_{PPv}V_{PPv}$, $\alpha = \frac{1}{V}(\alpha_{Pv}n_{Pv}V_{Pv} + \alpha_{PPv}n_{PPv}V_{PPv})$, and $C_P = n_{Pv}C_{P_{Pv}} + n_{PPv}C_{P_{PPv}}$. Aggregate moduli and velocities for single phase the two-phase aggregates are computed using the Voigt-Reuss-Hill (VRH) averaging scheme (Watt et al., 1976).

## 3 Results

### 3.1 Effects of Fe and Al alloying on the Pv-PPv phase boundary

The effects of Fe and Al alloying on the Pv-PPv phase boundaries are indicated in Fig. 1 for (a) $Mg_{1-x}Fe^{2+}_xSiO_3$, (b) $(Mg_{1-x}Fe^{3+}_x)(Si_{1-x}Fe^{3+}_x)O_3$, (c) $(Mg_{1-x}Al_x)(Si_{1-x}Al_x)O_3$, (d) $(Mg_{1-x}Fe^{3+}_x)(Si_{1-x}Al_x)O_3$ with the bearing element concentration $x = 0.10$. Solid lines are the boundaries obtained by taking the arithmetic average of the leftmost and rightmost boundaries obtained using LDA(+$U_{SC}$) and GGA(+$U_{SC}$) functionals; the gray shaded areas represent the uncertainties.

The blue-shaded areas are the regions where the Pv and PPv phases are most likely to coexist. The hazel and pink lines are adiabatic (Brown & Shankland, 1981) and superadiabatic geotherms (Boehler et al., 2008). All predicted transition pressure shifts trends agree with experiments on similar compositions (Catalli *et al*., 2009). The trends in widths of the two-phase regions, including the uncertainties, are also similar to those found by Catalli *et al.* (2009). The most striking differences are: a) the Pv-PPv coexistence region for the $Fe^{2+}$- bearing case obtained by Catalli *et al*. (2009) is shifted by 5-10 GPa to higher pressures; b) the discrepancy between calculations and experiments in Figure 1c) is due to the high $Al^{3+}$- concentration ($x = 0.25$) in the experiment (Tateno, 2005); c) the two-phase region in the ($Fe^{3+}$, $Al^{3+}$)-bearing system is somewhat narrower than in experiments. The two-phase region in the $Fe^{3+}$-bearing system (Fig. 1b) is in excellent agreement with Catalli *et al*. (2009). Except for the $Fe^{3+}$- bearing case in which the two phase region widens slightly with increasing temperature (Fig. 1b), in all other cases the two-phase coexistence regions widen at lower temperatures. The phase boundary and uncertainty in pure $MgSiO_3$ and the posperovskite mole fraction in the two phase regions showin in Fig. 1 are given in Figs. 2S and 3S. Numerical values for the two-phase boundaries are given in Table SI.

A fundamental aspect of the Pv-PPv transition in the alloys is the pressure range of the two-phase region along some reasonable geotherm, e.g., an adiabatic one (Brown & Shankland, 1981). Our predicted widths in these alloys are shown in Fig. 2a. The dark colors are the most likely Pv-PPv coexistence regions (blue areas in Fig. 1), and the lighter colors are the uncertainties originating in the choice of exchange-correlation functional in the DFT calculations (see Fig. S4 for more details in the *ab initio* uncertainty). $Fe^{2+}$ reduces the transition pressure, while $Al^{3+}$, $Fe^{3+}$, ($Fe^{3+}$, $Al^{3+}$)- increase it. These results obtained by systematic and consistent calculations offer consistent trends compared to measurements by Catalli et al. (2009). However, they contrast with results obtained in previous calculations by Wang & Tsuchiya (2019) that suggested the onset transition pressure should decrease in all cases.

Figs. 2b and 2c show widths inferred from experimental phase boundaries in these alloys (see Fig. 2 caption) and previous calculations (Wang et al., 2019), respectively. All measurements and calculations suggest a sizable pressure/depth range for these Pv-PPv transitions. Calculations offer narrower ranges, but DFT uncertainties are equally large. Table I summarizes the two-phase regions' widths along the Brown and Shankland geotherm and average Claeyron slopes shown in Fig. 1. Table SII gives the two-phase regions' widths obtained by others as well.

The Pv-PPv coexistence pressure range is known to vary with the alloying element concentration (Akber-Knutson et al., 2005; Andrault et al., 2010; Catalli et al., 2009; Tateno, 2005). This result is demonstrated in our calculations and in Fig. 3, where we plot only the average LDA(+$U_{sc}$)/GGA(+$U_{sc}$) two-phase region for concentrations $x = 0.08, 0.10$, and $0.12$. In all cases, the coexistence pressure range broadens with increasing concentration.

**3.2 Changes in seismic properties across the phase transition**

Changes in density ($\rho$), compressional ($V_P$), bulk ($V_\phi$), and shear ($V_S$) velocities across the phase transition depends on the width of the two-phase region, alloy concentration, and geotherm. Fig. 4 shows these properties' percentage changes with respect to the Pv property value as a function of pressure along adiabatic (Brown & Shankland, 1981) and superadiabatic (Boehler, 2000) geotherms for different alloys with $x = 0.10$. Each color depicts changes for a particular alloy. Shaded areas show the uncertainties that result from using LDA(+$U_{SC}$) and GGA(+$U_{SC}$) boundaries.

Overall, changes in seismic observables are within the magnitude of previous seismic studies (Sidorin et al., 1998; Wysession et al., 1998). All changes along the adiabatic geotherm are positive with the exception of $\Delta V_\phi(\%)$, which shows values of $\Delta V_\phi(\%) \sim -0.8\%$ in most cases (Figs. 4b*,g*,q*) and $\Delta V_\phi(\%) \sim -2.5\%$ for the $Al^{3+}$- case (Fig. 4l*). These negative values of $\Delta V_\phi(\%)$ reflect a decrease of the adiabatic bulk modulus ($K_S$) across the transition, in accordance with previous studies (Tsuchiya et al., 2004; Wentzcovitch et al., 2006; Wysession et al., 1998). In addition, $\Delta V_\phi(\%)$ displays a remarkable feature throughout two-phase regions. As a general rule, the high-pressure phase is denser. In the single-phase domains of these phases, smooth compression curves are expected. The two-phase region should have an anomalous compressibility caused by the change in the mole fractions of the phases involved (see Fig. S3), as previously pointed out in the case of the spin-crossover in iron in Fp (Wentzcovitch et al., 2009). The continuous increase in the amount of the denser high-pressure phase produces a softening anomaly in the isothermal, $K_T$, andin the adiabatic bul modulus, $K_S$. In the particular case of the Pv-PPv transition, the softening in $V_\phi$ is clearly visible, though the expected increase after the softening is less dramatic because the PPv phase is more compressible than the Pv phase right after the transition (Wentzcovitch et al., 2006). The Pv-PPv transition is an exception in this regard, but still displays such bulk modulus anomaly. Sometimes this bulk modulus softening might not be so evident, depending on the width of the two-phase region, the degree of equilibrium achieved in the system, and the differences in $K_S/K_T$ of the phases involved. However, this softening anomaly should be a general feature. Such softening should be expected in some of the elastic coefficients related to the bulk modulus, such as the compressional and off diagonal coefficients, as displayed across the spin crossover in iron in Fp (Wu *et al.*, 2013). A similar phenomenon has been clearly observed experimentally in phase transitions in the olivine system (Li & Weidner, 2008). However, the observed effect was time-dependent, also suggesting a kinetic contribution and the important of the rheological properties of the aggregates. Nevertheless, the observed bulk modulus softening is consistent with a component originating from the variation of phase abundances throughout the two-phase region.

Changes in shear velocity are relatively more significant, with values up to $\Delta V_S(\%) \sim 3.7\%$ for the $Fe^{3+}$- case (Fig. 4h*)). All other $\Delta V_S(\%)$ values are within $\sim 2.0$-$2.4\%$. This increase is consistent with seismic observations in some areas of the deep mantle (Sidorin *et al.*, 1998; Wysession et al., 1998) and previous calculations (Wentzcovitch *et al.*, 2006). The changes in compressional velocity, $\Delta V_P(\%)$, are positive in all cases with values varying between $\sim 0.3$-$0.9\%$ similar to the change in pure bridgmanite (Wentzcovitch et al., 2006). However, $\Delta V_P(\%)$ for $Fe^{2+}$ (Fig. 4a)) decreases between 95 GPa to 110 GPa. This distinct feature is caused by the softening in $K_S$ in the twophase region. Density changes, $\Delta \rho(\%)$, are $\sim 1.6\%$ for all cases. This observation is quite consistent with density discontinuities reported for the D" region (Sidorin et al., 1998; Wysession et al., 1998).

Positive and negative 'jumps,' also known as paired discontinuities, have also been reported (Thomas et al., 2004; Wysession et al., 1998). This phenomenon depends strongly on the geotherm and seems to correspond to a Pv-PPv-Pv sequence of transitions referred to as "double-crossing" (Hernlund et al., 2005; Hernlund, 2010). We also computed velocity and density changes along a superadiabatic geotherm (Boehler, 2000) (Figs. 4**). In this case, double-crossing occurs with PPv reverting to Pv, or almost so, for the $Fe^{3+}$-, $Al^{3+}$- and $Al^{3+}$, $Fe^{3+}$- cases, but not for the $Fe^{2+}$- case. The effect of the double-crossing on $\Delta V_P$ and $\Delta V_S$ is an initial increase/decrease followed by a reversal upon reduction of $n_{PPv}$ (e.g., most of Figs. 4**).

### 3.3 Effects on the adiabatic temperature gradient and on mantle aggregates

The nature and magnitude of the thermal boundary layer atop the core-mantle boundary (CMB) are still debatable. The Pv-PPv transition being exothermic contributes to the nonadiabatic nature of the temperature gradients in this region, with reported temperatures reaching up to 4,000 K. Conductive and to some extent, radiative heat transport across the CMB is the leading cause of the steep temperature gradient above the CMB; however, the Pv-PPv transition also contributes to the temperature gradient in this region. Fig. 5 shows the isentropic temperature profiles across this transition for all the alloys investigated. They are obtained by integrating the isentropic temperatures gradient (Eq. 5) starting from the same fixed point of the Brown & Shankland geotherm (T = 1873 K at 23 GPa). They differ from the latter by ~80 K at the CMB. They are ~1,300 K cooler than the Boehler geotherm (Boehler, 2000) since they do not include the effect of heat conduction across the CMB. None of the isentropes exhibit significant gradient changes across the Pv-PPv transition, as shown in the inset in Fig. 5. This is because both phases have very similar thermodynamic properties across the Pv-PPv transition (Tsuchiya et al., 2005). The procedure adopted here to compute the isentropic temperature profile is the same as that adopted in a recent study of the effect of the spin crossover in iron in Fp on the mantle geotherm (Valencia-Cardona et al., 2017).

### 4 Discussion

This study aims to clarify the effect of Fe and Al on the Pv-PPv phase transition and its potential consequences on seismic observables and temperature gradients in the deep mantle. Ultimately, a similar study must be carried out on aggregates in thermochemical equilibrium, but computations and experiments must be consistent first before one attempts to inspect detailed effects of aggregate composition on the Pv-PPv transition and shed light on remaining discrepancies between experiments on aggregates, e.g., with pyrolytic composition (Grocholski et al., 2012; Kuwayama et al., 2022). In this regard, the trends in our results are all consistent with measurements in bridgmanite with similar compositions (Catalli et al., 2009).

First, we verified that $Fe^{2+}$ decreases the transition pressure, an effect similar to that caused by a decrease in temperature, which increases the height of the D" discontinuity from the CMB. This behavior is opposite to that of ($Al^{3+}$), ($Fe^{3+}$) and ($Fe^{3+}$, $Al^{3+}$)- substitutions. Still, in all cases, the transition pressure shift is approximately proportional to the alloying element concentrations investigated, representing possible deep mantle bridgmanite compositions. In bridgmanite alone, the transitions are all quite broad and don't seem compatible with D" discontinuity of < 30 km (Lay, 2008). A change in composition across the D" discontinuity with a predominance of PPv on the high-pressure side seems a plausible scenario (Grocholski et al., 2012; Kuwayama et al., 2022). On face value, a superadiabatic geotherm could still produce double-crossing or, depending on the concentration, even prevent the phase transition from happening in ($Al^{3+}$), ($Fe^{3+}$) and ($Fe^{3+}$, $Al^{3+}$)- bearing $MgSiO_3$, but not in ($Fe^{2+}$)-bridgmanite in the absence of ferropericlase (Fp). "Double-crossing," as shown in most Figs. 4**, might not consist of two sharp transitions but a smooth change and reversal of velocities. It has been argued that in the presence of Fp, the change of iron partitioning throughout the transition increases the onset pressure of this transition and narrows the transition pressure range (Catalli et al., 2009). This effect is yet to be reproduced by *ab initio* calculations.

Second, it is believed that the exothermic nature of the Pv-PPv transition should contribute to the temperature gradient above the CMB (Nakagawa & Tackley, 2004). Regardless of the alloying element, there are no indications of a steep increase in the adiabatic temperature gradient throughout the two-phase region of the Pv-PPv phase transition. This is because the thermodynamic properties of both phases are very similar at the phase boundary (Tsuchiya et al., 2005). The steep temperature gradients in this region should be caused almost exclusively by conductive, and possibly radiative heat flows across the CMB.

Finally, we have also observed an anomalous softening of the bulk modulus throughout the two-phase coexisting region similar to that which occurs throughout the spin crossover in Fp (Wentzcovitch et al., 2009). This type of bulk modulus anomaly, however subtle it might be, is to be expected throughout two-phase regions since they result from the change in molar fractions of initial and final phases with different densities. A more complex time-dependent version of this phenomenon has been observed experimentally in the olivine system (Li & Weidner, 2008).

## 5 Conclusions

Using *ab initio* calculations, we examined the effects of Fe and Al on the Pv-PPv phase boundary and its consequences on the thermal structure of the deep mantle and seismic observables. Overall, the dependence of the phase boundaries on the composition agrees with experiments in bridgmanite with similar compositions (Catalli et al., 2009). The magnitude of overall velocity changes is also consistent with seismic observations. The onset pressures and widths of the two-phase regions change proportionally with alloying element concentrations with $Fe^{2+}$ decreasing the onset transition pressure and $Al^{3+}$, $Fe^{3+}$, and ($Fe^{3+}$, $Al^{3+}$) increasing it. We observe a softening of the bulk modulus in the Pv-PPv coexistence pressure range caused by the variable molar fraction of phases. A "double-crossing" of the phase boundary may cause only smooth changes/reversals in velocities but not sharp "lenses" for compositions in the range investigated here. In bridgmanite alone, the two-phase regions might be too broad to cause a < 30 km thick D" discontinuity (Lay, 2008). A change in composition across the D" discontinuity or throughout "double-crossings" with a predominance of PPv on the high-pressure side or inside the lenses seem a plausible scenario to be investigated in aggregates and compare with tomographic images. Finally, the changes in the adiabatic temperature gradients across the transition are minor in all cases. This transition should not contribute significantly to the thermal boundary layer in the D" region.

**Acknowledgments**

This work was supported by the National Science Foundation awards EAR-1918126 (JVC, GS, KS, and RMW) and EAR-2000850 (JZ and RMW) and in part by the US Department of Energy award DESC0019759 (RMW). This work used the Extreme Science and Engineering Discovery Environment (XSEDE), USA, which was supported by National Science Foundation, USA Grant Number ACI-1548562. Computations were performed on Stampede2, the flagship supercomputer at the Texas Advanced Computing Center (TACC), The University of Texas at Austin, generously funded by the National Science Foundation (NSF) through award ACI-1134872.

**Figures and Tables**

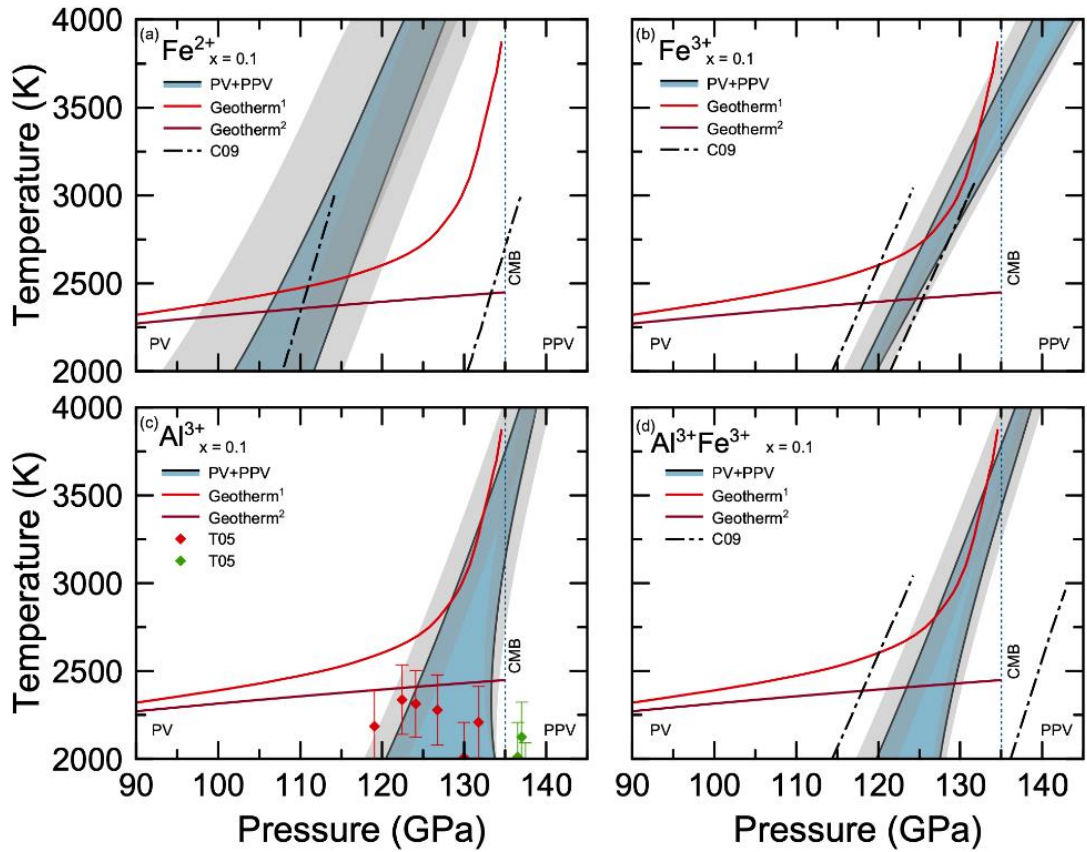

**Figure 1.** Phase boundaries of (a) $Mg_{1-x}Fe^{2+}_xSiO_3$, (b) $(Mg_{1-x}Fe^{3+}_x)(Si_{1-x}Fe^{3+}_x)O_3$, (c) $(Mg_{1-x}Al_x)(Si_{1-x}Al_x)O_3$, (d) $(Mg_{1-x}Fe^{3+}_x)(Si_{1-x}Al_x)O_3$, for a concentration x = 0.10. Solid blue lines are the boundaries. Gray shaded areas are the DFT related uncertain in the calculations. Blue shaded areas are the regions where the Pv and PPv phases coexist. Dashed lines and symbols are the experimental data from C09 (Catalli et al., 2009) and T05 (Tateno, 2005) respectively. Geotherm[1] and Geotherm[2] are those from Brown and Shankland (Brown & Shankland, 1981), and Boehler (Boehler, 2000), respectively. The vertical dashed-dotted line indicates the location of the core-mantle boundary (CMB).

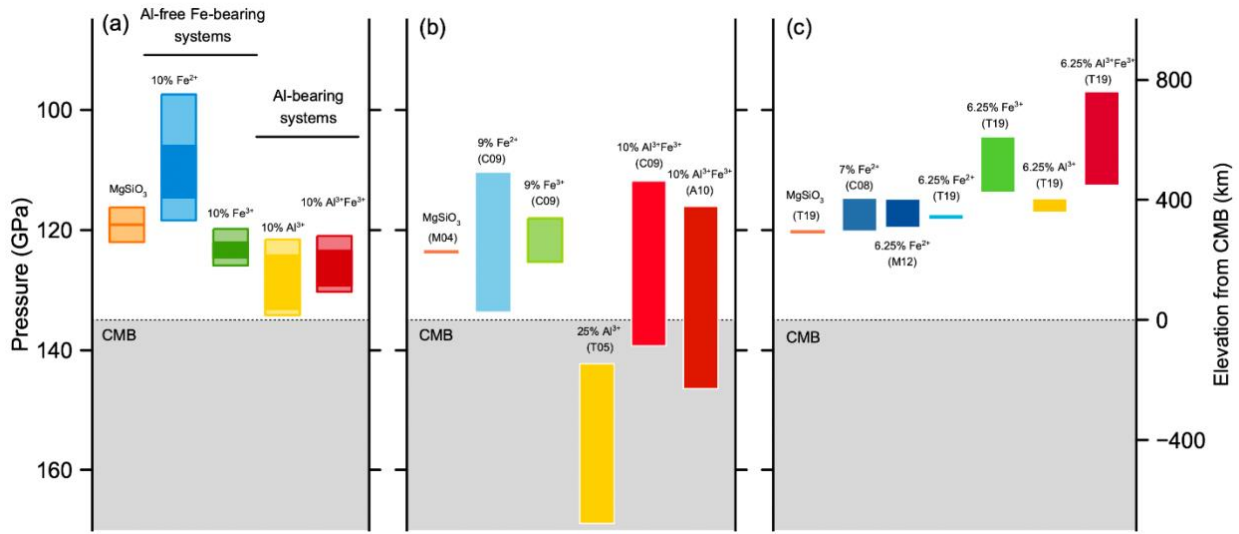

**Figure 2.** Depth and thickness of the Pv-PPv boundary along the Brown and Shankland geotherm (Brown & Shankland, 1981) for MgSiO$_3$ (orange), Mg$_{1-x}$Fe$^{2+}_x$SiO$_3$ (blue), (Mg$_{1-x}$Fe$^{3+}_x$)(Si$_{1-x}$Fe$^{3+}_x$)O$_3$ (green), (Mg$_{1-x}$Al$_x$)(Si$_{1-x}$Al$_x$)O$_3$ (yellow), (Mg$_{1-x}$Fe$^{3+}_x$)(Si$_{1-x}$Al$_x$)O$_3$ (red). The figure contains data reported in (a) this study for a concentration $x = 0.10$, (b) experimental (C09 (Catalli et al., 2009), T05 (Tateno, 2005), A10 (Andrault et al., 2010)) and (c) computational (C08 (Caracas & Cohen, 2008), M12 (Metsue & Tsuchiya, 2012), T19 (Wang et al., 2019)) studies for concentrations as labeled. The location of each colored bar represents the pressure and depth range where Pv and PPv phases coexist. In (a), the dark colored regions show the upper and lower bounds for the thickness obtained from the DFT calculations; the light-colored regions represent the DFT related uncertainties in the calculations of this work.

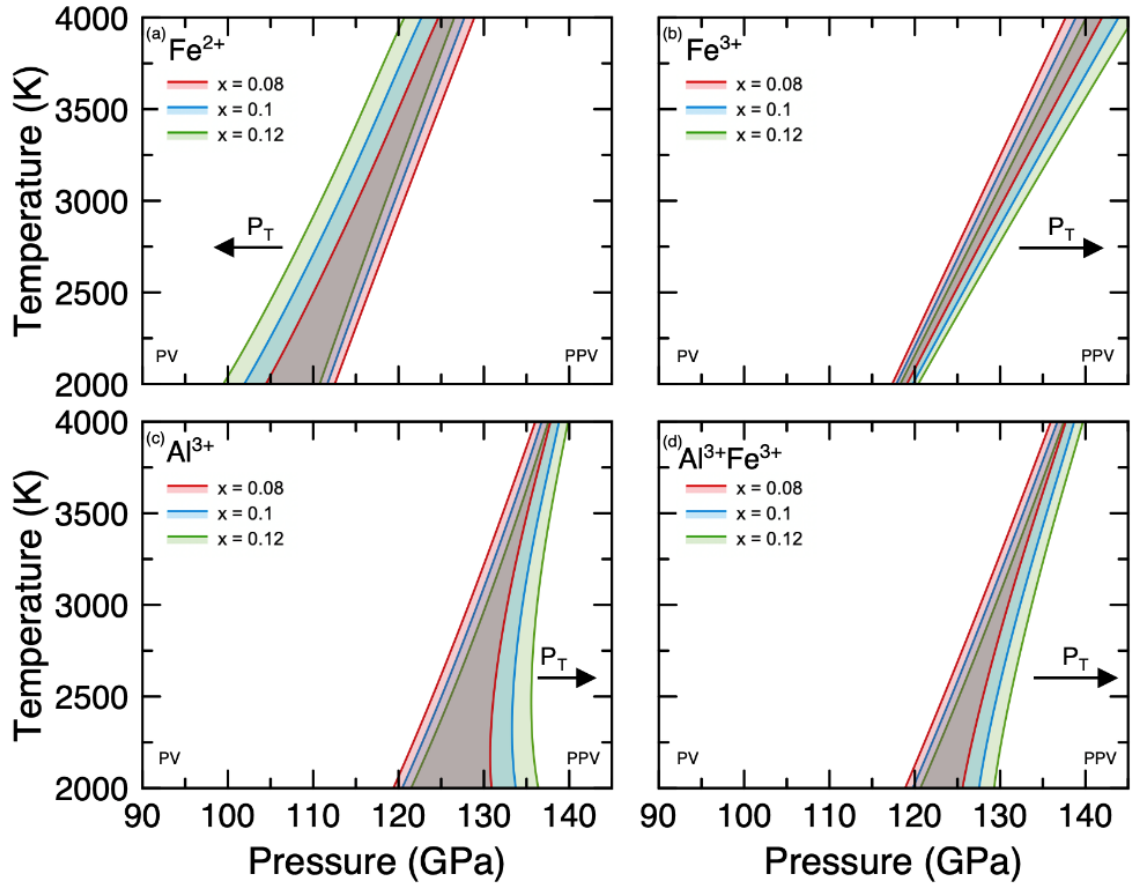

**Figure 3.** Phase boundaries of (a) $Mg_{1-x}Fe^{2+}_xSiO_3$, (b) $(Mg_{1-x}Fe^{3+}_x)(Si_{1-x}Fe^{3+}_x)O_3$, (c) $(Mg_{1-x}Al_x)(Si_{1-x}Al_x)O_3$, (d) $(Mg_{1-x}Fe3+x)(Si_{1-x}Al_x)O_3$, for a concentration x = 0.08 (red), 0.10 (blue), 0.12 (green).

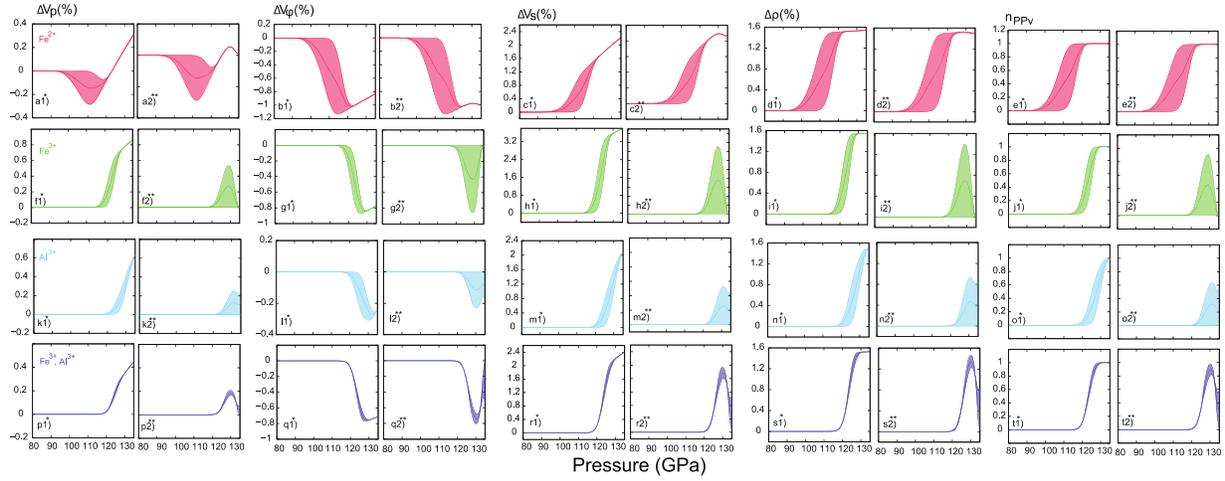

**Figure 4.** Calculated contrasts $\Delta V_P$ (%), $\Delta V_\varphi$(%), $\Delta V_S$ (%), $\Delta\rho$(%) and PPv fracion n in $Mg_{1-x}Fe^{2+}_xSiO_3$, $(Mg_{1-x}Fe^{3+}_x)(Si_{1-x}Fe^{3+}_x)O_3$, $(Mg_{1-x}Al_x)(Si_{1-x}Al_x)O_3$, $(Mg_{1-x}Fe^{3+}_x)(Si_{1-x}Al_x)O_3$ with x=0.1. The differences are calculated along the Brown & Shankland (Brown & Shankland, 1981) (*) and the Boehler (**) geotherms as: $\Delta M = 100 * (M_{PPv} - M_{PV})/M_{PV}$, with $M$ as: $V_P$, $V_\varphi$, $V_S$, and $\rho$. The shaded regions are limited by LDA(+$U_{sc}$) and GGA(+$U_{sc}$) results with the solid line in the middle being the average value.

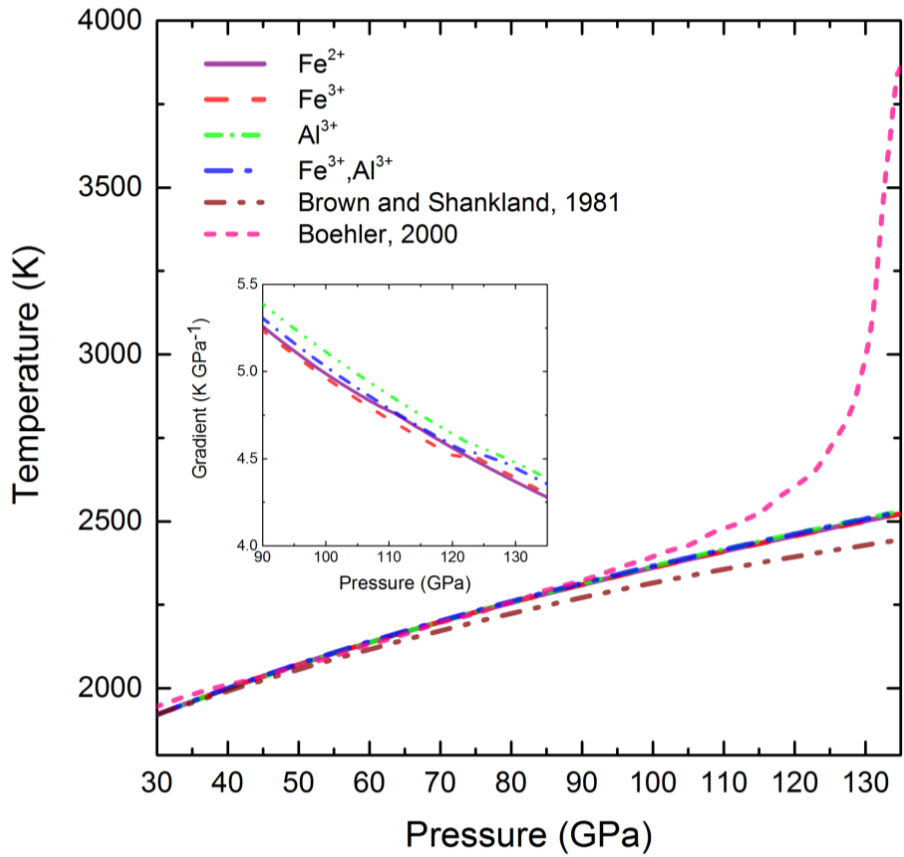

**Figure 5.** Adiabats of $Mg_{1-x}Fe^{2+}_xSiO_3$, $(Mg_{1-x}Fe^{3+}_x)(Si_{1-x}Fe^{3+}_x)O_3$, $(Mg_{1-x}Al_x)(Si_{1-x}Al_x)O_3$, $(Mg_{1-x}Fe^{3+}_x)(Si_{1-x}Al_x)O_3$ with x=0.1. The hazel/pink profiles correspond to that of Brown & Shankland (Brown & Shankland, 1981)/ Boehler (Boehler, 2000). The inset shows the temperature gradients across the transition.

**Table I.** Clapeyron slopes of the left (Left), right (R), and middle (M) of the two phase regions and widths of the two phase regions of along the Brown & Shankland (1981) geotherm for $Mg_{1-x}Fe^{2+}_xSiO_3$, $(Mg_{1-x}Fe^{3+}_x)(Si_{1-x}Fe^{3+}_x)O_3$, $(Mg_{1-x}Al_x)(Si_{1-x}Al_x)O_3$, $(Mg_{1-x}Fe^{3+}_x)(Si_{1-x}Al_x)O_3$ with $x = 0.1$. The slopes correspond to the linear high temperature range.

|  | $Fe^{2+}$ | $Fe^{3+}$ | $Al^{3+}$ | $Fe^{3+}, Al^{3+}$ |
|---|---|---|---|---|
| **L dP/dT (MPa/K)** | 10.2 ± 1.0 | 10.5 ± 0.1 | 8.1 ± 0.2 | 8.4 ± 0.3 |
| **R dP/dT (MPa/K)** | 8.2 ± 0.1 | 12.1 ± 0.3 | 3.2 ± 0.6 | 5.9 ± 0.2 |
| **M dP/dT (MPa/K)** | 9.1 ± 1.0 | 11.3 ± 0.8 | 5.6 ± 2.5 | 7.1 ±1.3 |
| **Width (GPa)** | 7.6 ± 6.2 | 2.5 ± 1.8 | 9.0 ± 1.8 | 5.8 ± 1.7 |
| **Width (km)** | 150 ± 125 | 50 ± 35 | 180 ± 35 | 120 ± 34 |